\documentclass[10pt,aps,pre,twocolumn,showkeys,superscriptaddress]{revtex4-1}
\usepackage{amsmath,amsfonts,amssymb}
\usepackage{color,physics,graphicx,lipsum}
\usepackage{hyperref,bm,comment,xcolor,soul}
\usepackage{enumitem}
\usepackage{pgfplots,tikz,pifont,pgfplots}
\usepackage{cleveref}
\usepackage[normalem]{ulem}
\hypersetup{colorlinks,linkcolor={red!50!black},citecolor={blue!90!black},urlcolor={blue!80!black}}

\newcommand{\ee}{\mathrm{e}}

\newcommand{\beginsupplement}{%
        \setcounter{table}{0}
        \renewcommand{\thetable}{S\arabic{table}}%
        \setcounter{figure}{0}
        \renewcommand{\thefigure}{S\arabic{figure}}%
        \renewcommand{\thesection}{S\arabic{section}}%
        \renewcommand{\theequation}{S\arabic{equation}}%
     }
     
\pgfplotsset{compat=1.18}
\pgfmathsetseed{5}%

\begin{document}

\title{On the area swept by a biased diffusion till its first-exit time: \\ Martingale approach and gambling opportunities}

\author{Yonathan Sarmiento}
\affiliation{ICTP -- The Abdus Salam International Centre for Theoretical Physics, Strada Costiera 11,  Trieste 34151, Italy}
\affiliation{International School for Advanced Studies (SISSA), Via Bonomea 265, Trieste 34136, Italy}

\author{Debraj Das}
\affiliation{ICTP -- The Abdus Salam International Centre for Theoretical Physics, Strada Costiera 11,  Trieste 34151, Italy}

\author{\'Edgar Rold\'an}
\email{edgar@ictp.it}
\affiliation{ICTP -- The Abdus Salam International Centre for Theoretical Physics, Strada Costiera 11,  Trieste 34151, Italy}


\begin{abstract}
Using martingale theory, we compute, in very few lines, exact analytical expressions for various first-exit-time statistics associated with one-dimensional biased diffusion. Examples include the distribution for the first-exit time from an interval, moments for the first-exit site, and functionals of the position, which involve memory and time integration. As a key example, we compute analytically the mean area swept by a biased diffusion until it escapes an interval that may be asymmetric and have arbitrary length. The mean area allows us to derive the hitherto unexplored cross-correlation function between the first-exit time and the first-exit site, which vanishes only for exit problems from symmetric intervals. As a colophon, we explore connections of our results with gambling, showing that betting on the time-integrated value of a losing game it is possible to design a strategy that leads to a net average win. 
\end{abstract}

\keywords{Martingale theory, First-passage processes, Brownian motion}
\maketitle


\section{Introduction}
\label{sec:intro}

The genesis of research in stochastic processes can be traced back to at least 1827, when Robert Brown~\cite{brown1828xxvii} observed the erratic movements of pollen grains suspended in water, later termed Brownian motion. The phenomenon remained unexplained until Einstein in 1905~\cite{einstein1905motion} developed a microscopic theory for the diffusive motion of Brownian particles, first tested in the lab by Perrin~\cite{perrin2013brownian}. Since then, Brownian motion, often referred to as diffusion, has become a key model for investigating diverse phenomena related to stochastic processes across various scientific disciplines, such as physics, chemistry, biology, and mathematics, as well as in economics, finance, and computer science. It has helped in many fundamental advancements of equilibrium and nonequilibrium statistical physics~\cite{kubo_brownian_1986,hanggi_introduction_2005}, fluid mechanics~\cite{Batchelor_1977,Mewis_Wagner_2011}, probability theory and stochastic differential equations~\cite{stratonovich1963topics,nelson1967dynamical,gardiner2004handbook}. Diffusion has also played a central role in studies of option pricing in finance~\cite{black1973pricing}, microrheology of viscoelastic materials~\cite{mason1995optical}, quantum fluctuations~\cite{gardiner2004quantum,campisi2011colloquium}, etc. One aspect of diffusion that has emerged as a topic of recent interest is the statistics of area-like functionals covered by a Brownian trajectory~\cite{majumdar_airy_2005,Kearney_2005,KearneyMJ2007,Kearney2014}. Specifically, such studies discover numerous practical uses in the fields of physics, mathematics, and computer science~\cite{majumdar2005BF}. Examples range from determining the expenses involved in building a table for data storage through linear probing with a random hashing algorithm~\cite{flajolet_analysis_1998}, analyzing statistics related to the maximal relative height of fluctuating interfaces~\cite{majumdar_airy_2005} to studying various discrete combinatorial problems found in graph theory that are associated with Bernoulli processes~\cite{Janson2007}.

In the last century, the development of the theory of stochastic processes has provided means to quantify a plethora of statistical properties of nonequilibrium processes driven by Brownian noise, such as that induced by thermal fluctuations. Such efforts paved the way, among other things, to the introduction of the mathematical foundations for what is now known today as martingales~\cite{doob}. Martingales, stochastic processes that serve as unbiased predictors of the future, become a key building block and also may provide a robust framework for understanding the dynamics of Brownian motion. In the context of nonequilibrium physics, martingales have found a recent revival leading to a novel `martingale' formalism for stochastic thermodynamics~\cite{roldan2022martingales}. This formalism has brought a plethora of exact results concerning various quantities, e.g., stopping-time and extreme-value statistics of thermodynamic quantities~\cite{roldan2022martingales}. One of the most important aspects of the martingale approach is that it makes the computation of averages of relevant quantities readily approachable. For example, by choosing an appropriate function, we may compute the mean area under the curve of a path that follows a Brownian motion with a directional bias. Previous works have explored this problem for a single absorbing boundary using techniques such as path-integral methods \cite{Kearney_2005, KearneyMJ2007, Janson2007, Abundo2013}.

Interesting recent papers about the stochastic area and universal first-passage statistics have provided insights into various aspects of stochastic processes. Studies on planar Brownian motion and linear ergodic diffusions have contributed to our understanding of stochastic areas, including the generation function, large deviation functions, and asymptotic properties~\cite{du2023large}. Additionally, the existence of anomalous scaling behavior has been found in the fluctuations of the area swept by one-dimensional Brownian motion, revealing deviations from traditional large deviations principles and uncovering singularities indicative of dynamical phase transitions~\cite{smith2023anomalous}. On the other hand, the theory of first-passage~\cite{redner2001guide,metzler_first-passage_2014}, which may also be realized as a sink of probability and be tackled by using probability flow analysis~\cite{sekimoto_derivation_2022}, has practical applications and relevance in various fields, e.g., in constraints arising in nonequilibrium thermodynamics~\cite{avanzini_methods_2023}, in the ranking of influential nodes in networks~\cite{bartolucci_ranking_2023}, etc. Furthermore, a bi-scaling theory has been proposed for first-passage times in confined compact processes, offering a comprehensive framework for understanding the probability density across different time scales~\cite{baravi2023first}.

In this work, we focus on the statistics of one-dimensional biased diffusion $X_t$ which follow the overdamped Langevin equation
\begin{align}
    \dot{X_t} = v + \sqrt{2D} \xi_t ,
    \label{eq:eom_Xit}
\end{align}
where $\xi_t=\dot{B}_t $ is the standard Gaussian white noise with zero mean mean $\expval{\xi_t} = 0$ and autocorrelation function $\expval{\xi_t \xi_{t'}} = \delta(t-t')$.
Here, $v$ is the drift velocity, which we will often take $v\geq 0$, and $D>0$ is the diffusion coefficient. We will assume that the initial position of the biased diffusion to be set at a fixed value $X_0$, which we will often take to be equal to zero, thus we have $X_t = X_0 + v t +{\sqrt{2 D}} B_t$ with $B_t$  being the  Wiener process.   

\begin{figure}[!htbp]
\centering
 \includegraphics[scale=0.5]{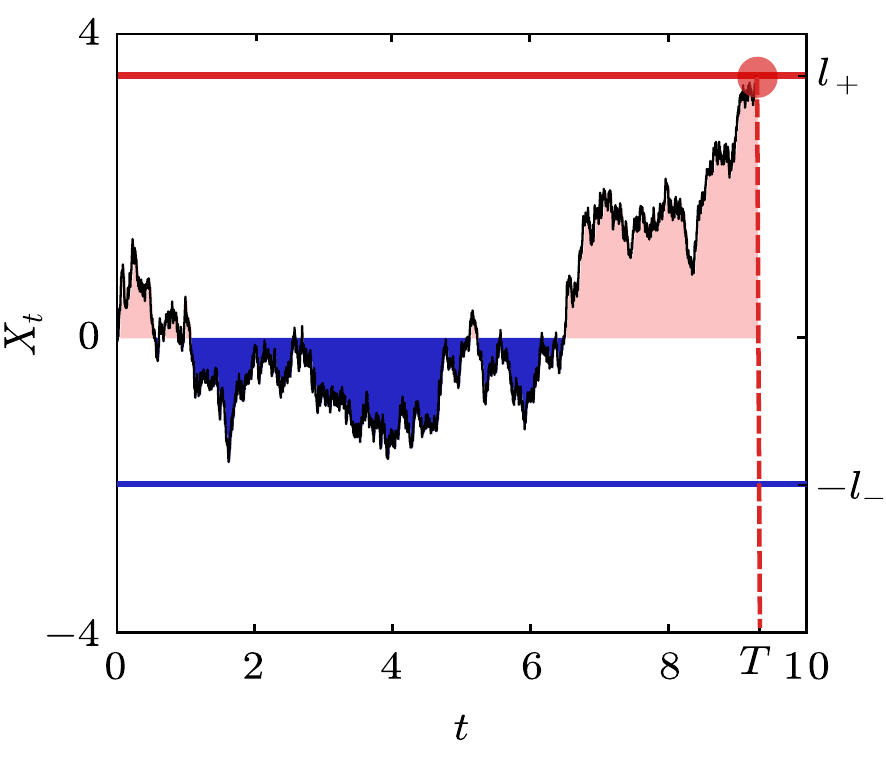}
\caption{Sketch of a biased diffusion $X_t$ given by Eq.~\eqref{eq:eom_Xit} with positive drift $v>0$.
The process stops at a stochastic time $T$ given by the earliest time  when the biased diffusion reaches either of two absorbing boundaries $l_{+}$ or $-l_-$. 
The black line depicts an example trajectory hitting the boundary $l_+$ at time $T$, which denotes the stopping time. The shaded colors illustrate the area swept by the process, with red color indicating positive contributions and blue color indicating negative ones.}
\label{fig1}
\end{figure}

We focus here on the dynamics~\eqref{eq:eom_Xit} within an interval~$[-l_-,l_+]$, with $l_- >0$ and $l_+>0$, provided that the diffusion starts within the interval, i.e., $X_0\in (-l_-,l_+) $. We investigate  first-exit-time statistics for the biased diffusion to first reach any of two absorbing boundaries located at $-l_-$ and $l_+$. In particular, we ask ourselves:
\begin{itemize}
    \item What is the probability density $P(T)$ for the first-exit time 
    \begin{equation}
        T = \text{inf} \qty{ t \geq 0 | X_t \notin ( -l_{-},l_{+} ) }
        \label{eq:exitttime}
    \end{equation}
     from the interval?
    \item What are the splitting probabilities $P_-$ and $P_+=1-P_-$ for the particle to first exit the interval from~$-l_-$ and from~$l_+$, respectively?
    \item Are the first-exit time $T$ and the first-exit position, i.e., the position at time $T$,~$X_T$ statistically correlated? Note that $X_T$ is a binary random variable that can assume a value equal to either $-l_-$ or $l_+$.
    \item Can one access the statistics of the   {\em area under the  stochastic curve} $X_t$ \begin{equation}
    A_t = \int_0^t \dd s \, X_s,\label{eq:area}
\end{equation}
  in the time interval $[0,t]$? Equation~\eqref{eq:area} gives   the net  area under the trajectory $X_{[0,t]}$ above the time axis. In other words, $A_t$  is a random variable  that  may also have negative values when the area swept by $X_t$ below zero exceeds the area swept above zero. For an illustration, see Fig.~\ref{fig1}, where the red and blue shaded regions under the trajectory $X_{[0,T]}$ show positive and negative contributions to $A_T$ with $T$ being the first-exit time.
\end{itemize}

Herein, we employ tools from the martingale theory to tackle all the aforementioned questions. First, we retrieve well-known analytical results in first-passage theory  through rather shorter  calculations than using traditional  approaches (e.g. path integrals, backward Fokker-Planck equations, image method, etc.). Moreover, we establish relations between moments and cross-correlations  of $T$ and $X_T$ which are valid for a broad class of {\em stopping times} which includes first-exit times as specific cases. Finally, we derive an analytical expression for the mean value of the area swept by a biased diffusion until its first-exit time from an arbitrary interval, which generalizes results from previous work~\cite{Kearney_2005}.

\section{Rudiments on martingale theory}

We serve here an appetizer on martingale theory, reviewing known results for the ease of  readers that may not be familiar with such concepts. We focus on a few well-known results that are applicable to our study, and refer readers to the recent treatise on ``Martingales for physicists"~\cite{roldan2022martingales} for further details. 

\subsection{Martingales, Submartingales and Doob's optional stopping theorem}
\label{sec:stop}
To warm up, we recall well-known statistical properties of diffusion. It is well known that a biased diffusion~\eqref{eq:eom_Xit} with positive drift  $v\geq 0$  is a {\em submartingale}, i.e., its  conditional average  obeys
\begin{equation}
    \langle X_t | X_{[0,s]}\rangle \geq X_s,
\end{equation}
which holds for any given past history $X_{[0,s]}\equiv \qty{X_{t'}}_{t' \in [0,s]}$ with $0 \leq s\leq t$. In other words, their average conditionally increases with time.  For unbiased diffusions $v=0$, one retrieves that the Wiener process is a {\em martingale}, i.e.,
\begin{equation}
    \langle B_t | B_{[0,s]}\rangle = B_s,
\end{equation}
for any given past history $B_{[0,s]}$ and $0\leq s\leq t$. In other words, unbiased diffusions represent ``fair games" which neither increase nor decrease conditionally with time. 

Central results in martingale theory are Doob's optional stopping theorems~\cite{doob} (see Ch.4 in~\cite{roldan2022martingales}). When applied to biased diffusion~\eqref{eq:eom_Xit} with $v\geq 0$, and fixed initial condition $X_0$, the following version of Doob's optional stopping theorem  for submartingales (OSTs) holds,
\begin{equation}
    \langle X_T \rangle \geq X_0.
    \label{eq:DoobX}
\end{equation}
Here, $T$ is a {\em stopping time} which is the first time $X_t$ satisfies a predefined criterion. The average $\langle X_T \rangle $ is done over the values that $X_t$ takes at time $T$ when the criterion is first met. Remarkably, the OSTs~\eqref{eq:DoobX} holds for a broad class of stopping times as long as two conditions are met:
\begin{enumerate}[label=\Roman*.]
    \item The long-time limit of the survival probability associated with $T$ vanishes, i.e., $P(T < \infty)=1$; and
    \item  $X_t$ is bounded at the stopping time, i.e., $\vert X_T \vert <\infty$.
\end{enumerate} 
The inequality~\eqref{eq:DoobX} indicates that all stopping rules obeying these two criteria lead to an average {\em gain} with respect to the initial value of the process. For unbiased diffusion $v=0$, Eq.~\eqref{eq:DoobX} becomes Doob's optional stopping theorem for martingales (OSTm)
 \begin{equation}
     \langle X_T \rangle = X_0,
     \label{eq:OSTm}
 \end{equation}
which holds for any martingale process $X_t$ with a fixed initial value $X_0$, see Ch.~4 in the Review~\cite{roldan2022martingales}. In other words,  martingales are {\em fair games} leading to no net gain or loss with respect to their initial value.

The Wiener process $B_t=X_t-vt$ illustrates that one can construct  a  martingale  from a submartingale. As we review in Sec.~\ref{sec:2b},  it is possible to systematically construct parametric families of martingales associated with submartingales. Such martingales are often written as functionals of $X_t$ and can be used, as we will show later, to extract  analytical expressions for various first-exit-time statistics  through  the OSTm~\eqref{eq:OSTm}. In particular, for biased diffusions,  the  first-exit time from a finite interval and the first-passage time to reach a single  threshold in the direction of the drift are two  suitable examples of~$T$, for which we can apply fruitfully the OSTm~\eqref{eq:OSTm} as we will show from Sec.~\ref{sec:3} and beyond.

\subsection{Constructing martingales from biased diffusions}
\label{sec:2b}

We refresh here readers about two celebrated parametric families of martingales that can be constructed using generic Markov processes as their building blocks. Particular emphasis will be given to specializing those families to the case of biased diffusion.

\underline{Dol\'{e}ans--Dade family.} 
The Dol\'{e}ans--Dade stochastic exponential~\cite{doleans1970quelques} associated with the Wiener process $ B_t$, defined as ${\mathcal{E}_t}(z) \equiv \exp[z B_t - {z^2}t/2] $, is a martingale for all $z\in\mathbb{R}$~\cite{protter_stochastic_2004,roldan2022martingales}, obeying $\langle \mathcal{E}_t(z)\vert B_{[0,s]}\rangle =\mathcal{E}_s(z)$ for $0 \leq s \leq t$. Such martingale, originally introduced by  Catherine  Doleans--Dade in the 1970s~\cite{doleans1970quelques},  is  also known as the zero-drift geometric Brownian motion. Substituting the Wiener process by $B_t = \qty(X_t -X_0 -v t ) /(\sqrt{2D})$  in the Dol\'{e}ans--Dade exponential, one retrieves that 
\begin{align}
   \label{eq:Nt_martingale}
    \mathcal{E}_t(z) = \exp\qty[ \frac{z}{\sqrt{2D}} \qty(X_t - X_0) - \qty(\frac{z v}{\sqrt{2D}}+ \frac{z^2}{2})t] 
\end{align}
provides a class of martingale processes associated with the biased diffusion~$X_t$. 
In other words, the stochastic exponential $\mathcal{E}_t(z)$ in Eq.~\eqref{eq:Nt_martingale}  is a martingale associated with the biased diffusion $X_t$ for all real values of the parameter $z\in \mathbb{R}$.

For the choice $z=-v\sqrt{2/D}$, we retrieve from Eq.~\eqref{eq:Nt_martingale} the thermodynamic martingale 
\begin{equation}
    \Sigma_t = \exp\left[-\frac{v}{D} (X_t-X_0) \right] ,\label{eq:MtXt}
\end{equation}
namely, that negative of the stochastic entropy production $S^{\rm tot}_t=(v/D) (X_t-X_0)$ associated with the biased diffusion is an exponential martingale in time-homogeneous stationary states~\cite{chetrite2011two,neri2017statistics}. Such result is the cornerstone of the martingale theory for stochastic thermodynamics in nonequilibrium stationary states, see~\cite{roldan2022martingales}. As we will see in Sec.~\ref{sec:3} and beyond, other choices for $z$ in the Dol\'{e}ans--Dade family are instrumental for first-passage-time calculations.

\underline{Dynkin family.}
Eugene  Dynkin~\cite{dynkin2000selected} proved rigorously that one can systematically construct martingales associated with a generic Markov process and a function $g \equiv g_t(x)$ that may depend explicitly on $t$ and $x$. 
In particular,
\begin{equation}
    \mathcal{Y}_t [g] = g_t(X_t) - g_0(X_0) - \int_0^t \dd s [\partial_s g_s(X_s) + \mathcal{L} g_s (X_s)]
    \label{eq:Dynkin's martingales}
\end{equation}
constitutes a martingale with respect to $X_t$ for any family of real-valued, smooth bounded function $g_t(x)$, for which the derivatives $\partial_t g_t(x)$, $\partial_x g_t(x)$, and $\partial^2_x g_t(x)$ exist~\cite{roldan2022martingales}. Here $\mathcal{L}$ denotes the generator of the Markov process $X_t$ at time $t$, for which  its instantaneous density $\rho_t(x)$ follows $\partial_t \rho_t(x) = \mathcal{L}^{\dag}  \rho_t(x)$, where $\mathcal{L}^{\dag}$ is the adjoint of the operator $\mathcal{L}$. When specialized over biased diffusion (Eq.~\eqref{eq:eom_Xit}), we have $\mathcal{L^{\dag}}= -v\partial_x + D\partial_x^2$ and $\mathcal{L}=v\partial_x + D\partial_x^2$, and therefore, Dynkin's martingales read
\begin{eqnarray}
    \mathcal{Y}_t [g] &=& g_t(X_t) - g_0(X_0) \nonumber\\
    &-& \int_0^t \dd s \left[ \left ( \partial_s g_s + v \partial_x g_s + D\partial_x^2 g_s \right) (X_s)\right],
    \label{eq:Dynkin's Langevin}
\end{eqnarray}
and obey $\langle\mathcal{Y}_t[g]\vert X_{[0,s]}\rangle=\mathcal{Y}_s[g]$ for any $t\geq s\geq 0$.

\subsection{Generic properties of stopping times for drift-diffusion dynamics }
\label{sec:ddd}

Expanding the Dol\'{e}ans--Dade exponential associated with $B_t$ around $z$ small, one may write 
\begin{align}
    \mathcal{E}_t( z ) 
    &= 1 + z B_t + \frac{z^2}{2!} (B^2_t-t)+\frac{z^3}{3!} (B^3_t-3tB_t) + \ldots,\nonumber\label{eq:doleansexp}\\
    &=\sum_{n=0}^{\infty} \frac{z^n}{n!} H^{(n)}_t, 
\end{align}
 and show that  the stochastic processes  in the expansion~\eqref{eq:doleansexp}, which we denote by $H^{(n)}_t $, are  all martingales satisfying $\langle H^{(n)}_t \vert B_{[0,s]}\rangle  = H^{(n)}_s$ for all $t\geq s\geq 0 $ and for all $n\geq 0$~\cite{roldan2022martingales}. Here, the martingales $H_t^{(n)}$ can be written in the form of Hermite polynomials evaluated over realizations of the Wiener process as
\begin{equation}
    H^{(n)}_t = (-t)^n  \left.\exp\left(\frac{B^2}{2t}\right) \frac{\mathrm{d}}{{\mathrm{d}} B^n} \exp\left(-\frac{B^2}{2t}\right)\right|_{B=B_t}.
\end{equation}

Considering the $n=1$ martingale  defined with respect to the drift-diffusion process, we may write $H_t^{(1)} = B_t= \qty(X_t -X_0 -v t ) /(\sqrt{2D})$.
Then, applying Doob's OSTm~\eqref{eq:OSTm} to the martingale $H_t^{(1)}$, i.e., $ \expval{H_T^{(1)}}=H_0^{(1)}=0 $, we get $\expval{ (X_T - X_0  - v T)}/ (\sqrt{2D}) =  H_0^{(1)} =0$, which leads to  the mean stopping time:
\begin{align}
     \expval{T} &= \frac{\expval{X_T} -  X_0 }{v} . \label{eq:MET-gen-x0}
\end{align}
Similarly, applying the OSTm~\eqref{eq:OSTm} to higher-order martingales $H_t^{(n)}$, one finds the  $(n\geq 2)$-th moments of the stopping time for the drift-diffusion process. For example,  we obtain from  $H_t^{(2)}=B_t^2-t$ and $H_t^{(3)}=B_t^3-3t B_t$ that
\begin{align}
    \expval{T^2} =  \frac{X_0^2 - \expval{X_T^2} }{v^2} + \frac{2  \expval{X_T T}}{v} + \frac{2D \expval{T}}{v^2} ,
    \label{eq:MET2-gen}
\end{align}
and
\begin{eqnarray}
    \expval{T^3} &=& \frac{\expval{X_T^3} - X_0^3}{v^3} - \frac{3}{v^2}  \expval{X_T^2 T} + \frac{3}{v} \expval{ X_T T^2 }\nonumber \\  &-&  \frac{3  X_0}{v^3} \expval{X^2_T} +\frac{3}{v^2} \left(2 D - v X_0\right)\expval{T^2}     \nonumber\\
    &+&   \frac{6}{v^3} \left( v X_0 - D \right) \expval{X_T T} + \frac{3  X_0^2}{v^3} \expval{X_T} \nonumber\\ &+& \frac{3X_0}{v^3} \left(2 D - v X_0  \right)\expval{T}   ,
     \label{eq:MET3-gen}
\end{eqnarray}
respectively.
Equations~\eqref{eq:MET2-gen} and~\eqref{eq:MET3-gen} reveal  that computing $(n>1)-$th moment of the stopping time $T$ requires {\em in general} the knowledge of the cross correlations $\expval{ X^{m}_T T^{n}}$.

\subsection{First-exit-site distribution for asymmetric intervals}
\label{sec:fes-asymm}

For the most generic case with two absorbing boundaries at $-l_{-}$ and $l_{+}$ with $l_{\pm}> 0$,  the stopping time $T$ is given by the first-exit time from the interval $(-l_{-},l_{+})$, i.e., one has
$T = \text{inf} \qty{ t \geq 0 | X_t \notin ( -l_{-},l_{+} ) }$.
Consequently, at the stopping time $T$, the position of the particle $X_T$ becomes a binary random variable that takes two possible values $-l_{-}$ and $l_{+}$, and thus the first-exit-site distribution is given by
\begin{equation}
\label{eq:P(X_T)_general}
    P(X_T) = P_+ \delta_{X_T,l_+} + P_- \delta_{X_T,-l_-},
\end{equation}
where $\delta_{i,j}$ denotes Kronecker's delta. 
We thus denote in the following the probabilities of $X_T$ taking on the values $-l_{-}$ and $l_{+}$ by $P_{-} \equiv P(X_T = -l_{-})$ and $P_{+} \equiv P(X_T = l_{+})$, respectively. 

Applying Doob's optional stopping theorem to the martingale $\Sigma_t$ defined in Eq.~\eqref{eq:MtXt},  we get $   \expval{\exp(\Sigma_T)} =1$, i.e.,
 \begin{equation}
    \underbrace{\expval{\exp\left[-\frac{v}{D} X_T \right]}}_{\displaystyle P_+ \exp(-vl_+/D) + P_- \exp(vl_-/D)}= \exp\left[ - \frac{v X_0}{D} \right] ,\label{eq:DoobPp}
\end{equation}
which together with the normalization condition $P_+ + P_- =1$ yields
\begin{eqnarray}
    P_{-}  &=& \frac{ \exp[ - v (l_{-} + X_0) / D ] - \exp[ - v (l_{-} + l_{+}) / D ] }{1- \exp[- v (l_{-} + l_{+}) / D ]  }, \nonumber \\
    P_{+}  &=& \frac{  1- \exp[- v (l_{-} + X_0) / D ]  }{1- \exp[ - v (l_{-} + l_{+}) / D ]  }  \, .\label{eq:P+P-assym_FB_x0}
\end{eqnarray}
The $m-$th moment of $X_T$ is given by 
\begin{equation}\label{eq:20}
    \expval{X^m_T} =  (- l_{-})^m P_{-} + l^m_+ P_{+}.
\end{equation} 
Substituting Eq.~\eqref{eq:20} for $m=1$ in Eq.~\eqref{eq:MET-gen-x0}  we find that the mean exit time reads
\begin{equation}\label{eq:21}
    \expval{T} = \frac{l_{+} P_{+}  - l_{-} P_{-} -X_0}{v}.
\end{equation}  Similarly, the $n-$th moment of the first-exit time $T$ is given by 
\begin{equation}\label{eq:22}
    \expval{T^n} = P_{-} \expval{T^n| X_T = -l_{-}} + P_{+} \expval{T^n|X_T= l_{+}} ,
\end{equation}
where the moments $\expval{T^n| X_T = -l_{-}}$ and $\expval{T^n|X_T= l_{+}}$ are to be computed with respect to the conditional probabilities $P(T|X_T = - l_-)$ and $P(T|X_T =  l_+)$, respectively.
Furthermore, the cross-correlations between the exit time and the exit site read
\begin{eqnarray}\label{eq:23}
    \expval{X_T^m T^n}  &=& P_{-} \, (-l_{-})^m  \, \expval{T^n| X_T = -l_{-}}  \\
    &+&   P_{+}  \, l_{+}^m \, \expval{T^n| X_T = l_{+}}.\nonumber
\end{eqnarray}
We remark that $X_T$ and $T$ are in general correlated random variables, which implies that it is highly non-trivial to compute $\expval{T^n}$ and $\expval{X_T^m T^n} $ through Eqs.~\eqref{eq:22} and~\eqref{eq:23}.  As we will show in the next Sec.~\ref{sec:symmetry}, such calculations can be enormously simplified for the specific case of a exit-time problem from a symmetric interval around the origin.

\subsection{First-exit-time symmetry for symmetric intervals}
\label{sec:symmetry}

For symmetric thresholds $l_+=l_-=l$ about an initial condition $X_0 = 0$, it was shown that the following first-passage-time symmetry 
\begin{equation}
    P(T|X_T=l)=P(T|X_T=-l),
    \label{eq:FPTsym}
\end{equation}
holds~\cite{roldan2015decision,krapivsky2018first,roldan2022martingales,dorpinghaus2018testing,dorpinghaus2023optimal}. Equation~\eqref{eq:FPTsym} implies that in case of symmetric thresholds the statistics of the trajectories that reach the positive threshold $l$ is the same as that of the trajectories that reach the negative threshold $-l$. We remark that this result holds even in presence of a non-zero drift. In other words, one may say that for symmetric thresholds even with $v\neq 0$ the distributions of the first-passage time for trajectories reaching either $l$ or $-l$ are the same. Consequently, the first-passage  symmetry~\eqref{eq:FPTsym} also implies that the conditional density of the escape time obeys
\begin{equation}
    P(T\vert X_T) = P(T).
\end{equation}
Therefore, for symmetric intervals, the mutual information between the exit time and the exit site vanishes
\begin{equation}
    I(X_T:T)=\left\langle  \log \frac{P(T,X_T)}{P(T)P(X_T)}\right\rangle =0,\label{eq:optI}
\end{equation}
thus the random variables  $X_T$ and $T$ are uncorrelated. Equation~\eqref{eq:optI} was mathematically proven in Ref.~\cite{dorpinghaus2018testing} in context of binary sequential hypothesis testing  with symmetric error probabilities. More precisely, Corollary~2 in~\cite{dorpinghaus2018testing} implies Eq.~\eqref{eq:optI} for the specific example of a binary decision among two hypotheses: $\mathsf{H}_0$ the dynamics is governed by $\dot{X}_t=v+\sqrt{2D}\xi_t$, and  $\mathsf{H}_1$ the dynamics is governed by $\dot{X}_t=-v+\sqrt{2D}\xi_t$, see also~\cite{dorpinghaus2023optimal} for further details. 

Interestingly, condition~\eqref{eq:optI} implies  that  the cross-correlation function  between the exit site and the exit time vanishes, i.e.,
\begin{equation}
\label{eq:symm-cross-corr}
    C(X_T,T) =\expval{X_T T}-\expval{X_T}\expval{T}  =0,
\end{equation}
and similarly the cross-correlation function of arbitrary powers is given by
\begin{equation}
    C(X_T^m,T^n) =\expval{X_T^m T^n}-\expval{X_T^m}\expval{T^n}  =0,\label{eq:momentsindepe}
\end{equation}
for any integer $m$ and $n$. Equation~\eqref{eq:momentsindepe} is instrumental to obtain analytical expressions for the $n-$th moment of the first-exit time from a symmetric interval around the origin. This can be easily grasped by noting that for symmetric intervals with $X_0=0$ one has (see Eqs.~\eqref{eq:20} and~\eqref{eq:21})
\begin{align}
    \expval{X^m_T} = \begin{cases}
        l^m \quad \quad  &\text{for $m$ even}, \\
         l^{m-1}  v \expval{T}   \quad \quad  &\text{for $m$ odd}, 
    \end{cases}
    \label{eq:XmT-symm-FB}
\end{align}
and then by using $\expval{X_T^m T^n}=\expval{X_T^m}\expval{T^n}$ in  Eqs.~\eqref{eq:MET2-gen}--\eqref{eq:MET3-gen}, which reveals  that  the $n-$th  moments (with arbitrary $n>1$) of the exit time can be expressed in terms of simple functions of solely the first moment~$\expval{T}$.

\section{First-passage statistics}
\label{sec:3}

\subsection{Single absorbing boundary}
\label{sec:sab}
Let us now choose a stopping criterion for the process $X_t$.
Suppose the process starts at $X_0 =0$, there is an absorbing boundary at $l>0$, and the process stops as soon as it  reaches the absorbing boundary. The stopping time $T$ in this case is defined by
\begin{align}
    T \equiv \text{inf}\qty{t\geq 0 | X_t \geq l } ,
    \label{eq:T-def-single-boundary}
\end{align}
which is nothing but the first-passage time to  the absorbing boundary $l$. The first-passage properties of a drift-diffusion process are well-studied in the literature. Arguably, the method of images may be considered as one of the simplest approach to obtain closed-form solutions for the  probability density $P(T)$ of the stopping time $T$ and survival probability $S(t)$ at a given time $t$~\cite{redner2001guide}. However, in the following, we show how one may obtain the results in an elegant way by using the Martingale theory. This method often referred to as the Gerber--Shiu technique helps in computing the Laplace transforms of probability densities of stopping times using martingale properties and has been used in mathematical finance~\cite{lin1998double}, mathematical psychology~\cite{srivastava2017martingale}, stochastic thermodynamics~\cite{neri2022universal}, etc.

In order to obtain the density $P(T)$, we recall that the Dol\'eans-Dade exponential  $\mathcal{E}_t(z)$ defined in Eq.~\eqref{eq:Nt_martingale} is a martingale and satisfies $\mathcal{E}_0 = 1$.
Then, applying Doob's OSTm~\eqref{eq:OSTm} to this martingale,  we obtain $\expval{\mathcal{E}_T} = \mathcal{E}_0 =1$, which implies the exact relation
\begin{align}
\label{eq:Et_avg}
   \expval{ \exp\qty[   -\underbrace{ \qty(\frac{z v}{\sqrt{2D}}+ \frac{z^2}{2})}_{\displaystyle s} T] } =  \exp\qty(  -\frac{z l}{\sqrt{2D}} ) \, ,
\end{align}
 where we have used the fact that $X_T=l$. 
 Equation~\eqref{eq:Et_avg} provides a direct route to the Laplace transform of the first-exit-time probability density  $P(T)$ as $\widetilde{P}(s) \equiv {\mathfrak{L}} [P(T)] (s)= \langle\exp(-sT)\rangle =\int_{0}^{\infty} \dd{T} \exp(-s T) P(T)$ with $s>0$. To obtain the time-dependent probability density, we reparametrize the argument of the exponential in the left-hand-side of Eq.~\eqref{eq:Et_avg} using the relation $s(z) =  {z v} /
 \sqrt{2D} + {z^2}/{2} $, and compute its roots 
$z_{\pm}(s) \equiv \qty(-v \pm \sqrt{v^2 + 4 D s} ) /(\sqrt{2 D})$, from which only the positive root $z_{+}$ gives us $s>0$ for $v>0$. By specializing $z$ in to the positive root, i.e., substituting 
\begin{equation}
    z=\frac{-v + \sqrt{v^2 + 4 D s} }{\sqrt{2 D}},
    \label{eq:zval}
\end{equation} in the right-hand side of Eq.~\eqref{eq:Et_avg}, 
one identifies its left-hand side as the Laplace transform of the first-exit-time distribution, which reads 
\begin{eqnarray}
    \widetilde{P}(s) = \exp\qty(  \frac{v l - l\sqrt{v^2 + 4 D s} }{2 D}   ).
    \label{eq:Ptilde_single}
\end{eqnarray}
Performing the inverse Laplace transform we obtain from Eq.~\eqref{eq:Ptilde_single} the  probability density of the stopping time~\cite{redner2001guide}
\begin{eqnarray}
P(T) &= & \mathfrak{L}^{-1}[\widetilde{P}(s)](T) \nonumber \\
&=& \frac{l}{ \sqrt{ 4 \pi D T^3}} \exp[-\frac{(l - v T)^2}{4 D T} ] .
\end{eqnarray}
which was also obtained in textbook reference with, e.g., the image method~\cite{redner2001guide} following rather longer calculations.

\subsection{Two symmetric absorbing boundaries}

We now consider the case with two absorbing boundaries at $-l$ and $l$ symmetrically placed about the initial position~$X_0=0$. 
Here, the process $X_t$ stops as soon as it reaches either of the absorbing boundaries.
Consequently, the first-exit time~\eqref{eq:exitttime} from the interval $(-l,l)$  in this case is given by
\begin{align}
    T = \text{inf} \qty{ t \geq 0 | X_t \notin (-l,l) } \, .
    \label{eq:T-def}
\end{align} 
Unlike the case of a  single absorbing boundary, here the exit site $X_T=\{l,-l\}$ becomes a binary random variable with probability distribution $P(X_T)=P_+\delta_{X_T,l} + P_-\delta_{X_T,-l}$, where $P_+$ and $P_-=1-P_+$ are the absorption probabilities at sites $l$ and $-l$, respectively. Despite this additional complexity, we can crack key first-exit-time statistics in few lines using martingales.

To derive the density $P(T)$ of the first-exit time~\eqref{eq:T-def}, we follow analogous  steps as in the previous Sec.~\ref{sec:sab}. The first-exit time~\eqref{eq:T-def} is yet another example of stopping time  for which  Doob's OSTm~\eqref{eq:OSTm} applies. Specializing again  the OSTm~\eqref{eq:OSTm} to  the Dol\'eans-Dade exponential  $\mathcal{E}_t(z)$ defined in Eq.~\eqref{eq:Nt_martingale}, we obtain $\expval{\mathcal{E}_T} = \mathcal{E}_0 =1$, which in this case reads
\begin{align}
\label{eq:26}
   \expval{ \exp\qty[ \frac{zX_T}{\sqrt{2D}} - \qty(\frac{z v}{\sqrt{2D}}+ \frac{z^2}{2}) T] } = 1 \, .
\end{align}
Evaluating the average on the left-hand side of Eq.~\eqref{eq:26} is far from straightforward for generic exit times, as it requires knowledge on the joint distribution $P(X_T,T)$. However, for the first-exit time from a symmetric interval  the exit site $X_T$ and first-exit time $T$ are statistically independent [see Sec.~\ref{sec:symmetry}]. As a result, the left-hand side of Eq.~\eqref{eq:26} factorizes
\begin{eqnarray}
\label{eq:27}
   &&\expval{ \exp\qty[ \frac{zX_T}{\sqrt{2D}} - \qty(\frac{z v}{\sqrt{2D}}+ \frac{z^2}{2}) T] } \nonumber\\
   &&=\expval{ \exp\qty[ \frac{zX_T}{\sqrt{2D}} ] } \expval{- \qty(\frac{z v}{\sqrt{2D}}+ \frac{z^2}{2}) T}.
\end{eqnarray}
In the right-hand side of Eq.~\eqref{eq:27}, the first average is done with respect to the first-exit-site distribution $P(X_T)$, whereas the second average is to be computed with respect to the first-exit-time distribution $P(T)$. 
Combining Eq.~\eqref{eq:26} with Eq.~\eqref{eq:27}, we obtain
\begin{equation}
\label{eq:ave_E_t_symetric}
       \expval{ \exp\qty[- \qty(\frac{z v}{\sqrt{2D}}+ \frac{z^2}{2}) T] } = \expval{ \exp \left( \frac{z X_T}{\sqrt{2D}}  \right) }^{-1} .
\end{equation}
Specializing $z =  \qty(-v + \sqrt{v^2 + 4 D s} ) /(\sqrt{2 D})$ [See Eq.~\eqref{eq:zval}], we obtain that the  Laplace transform $\widetilde{P}(s)$ of the first-exit-time probability density $P(T)$, i.e, 
\begin{align}
\widetilde{P}(s) =  \expval{\exp\qty( \frac{z X_T}{\sqrt{2D}} )}^{-1}.
\label{eq:Laplace_P(T)_sim}
\end{align}
The average of the exponential function on the right hand side of Eq.~\eqref{eq:Laplace_P(T)_sim} may be obtained by using Eq.~\eqref{eq:P+P-assym_FB_x0} in Eq.~\eqref{eq:P(X_T)_general} with $l_+ = l_- = l$. Then, and after some algebra  Eq.~\eqref{eq:Laplace_P(T)_sim} reads
\begin{equation}
\label{eq:Laplace_simm}
    \widetilde{P}(s) =  \cosh \left(\frac{l v}{2 D}\right) \text{sech}\left(\frac{l \sqrt{4 D s+v^2}}{2 D}\right) ,
\end{equation}
which in the limit $s\to 0$, yields $\lim_{s\to 0} \widetilde{P}(s) = 1$, and thus shows the normalization of $P(T)$.  Noting that the inverse Laplace transform of $\sech(b \sqrt{s})$ is given by~\cite{roberts_table_1966,oberhettinger_tables_1973}
\begin{align}
    {\mathfrak{L}}^{-1}[\sech(b\sqrt{s})](t) = & \frac{ b}{\sqrt{4 \pi t^3}} \sum_{n=-\infty}^{\infty}  (-1)^n (1-2n) \nonumber \\
    & \times \exp\left[-\frac{b^2(2n-1)^2}{4 t}\right] , 
\end{align}
one may invert Eq.~\eqref{eq:Laplace_simm} to obtain the exit-time distribution $P(T) = {\mathfrak{L}^{-1}} [\widetilde{P}(s)](T)$ as
\begin{align}
\label{eq:PT_symm}
    P(T) =  & \frac{l   }{ \sqrt{4 \pi D T^3} } 
 \cosh\qty(\frac{lv}{2D}) \exp\qty(-\frac{v^2 T}{4D}) \nonumber \\
 & \times \sum _{n=-\infty}^{\infty } (-1)^n (1-2n) \exp\left[-\frac{l^2  (2n-1)^2}{4 D T} \right] \, .
\end{align}
For an alternative expression of $P(T)$ we refer the reader to Eq.~(13) in ref.~\cite{srivastava2017martingale}. We note that in refs.~\cite{roberts_table_1966,oberhettinger_tables_1973}, the inverse Laplace transform of $\sech(b \sqrt{s})$ is mistakenly given by $- (1/b^2) \qty[ \partial_\nu \theta_{1}\qty(\nu/2| t/b^2) ]_{\nu = 0}$, where the Theta function $\theta_1(\nu|x)$ is  defined by~\cite{roberts_table_1966}
\begin{equation}
\theta_1(\nu|x) \equiv \frac{1}{\sqrt{\pi x} } \sum_{n=-\infty}^{\infty} (-1)^n \exp[-\frac{1}{x} \qty(\nu +n - \frac{1}{2})^2] .
 \end{equation}
However, the correct expression for the inverse Laplace transform of $\sech(b \sqrt{s})$ reads $ (1/b^2) \qty[ \partial_\nu \theta_{1}\qty(\nu/2| t/b^2) ]_{\nu = 0}$.
The $n$-th moment of the exit time $T$ may be written in terms of $\widetilde{P}(s)$ as
\begin{equation}
     \langle T^n \rangle = (-1)^n \frac{\dd^n}{\dd s^n}\widetilde{P}(s) \Big|_{s=0} ,
\end{equation}
using which along with Eq.~\eqref{eq:Laplace_simm}, the first three moments are obtained as
\begin{align}
\label{eq:Texp-sym}
   \expval{T} &= \frac{l}{v} \tanh\qty(\frac{v l}{2 D} )   \,  , \\
    \expval{T^2} &=  2 \expval{T}^2 + \frac{2D}{v^2} \expval{T}  -\frac{l^2}{v^2} ,
    \label{eq:T2exp-sym} \\
    \expval{T^3} &= 6 \expval{T}^3  + \frac{12 D}{v^2} \expval{T}^2 +\qty( \frac{12 D^2 }{v^4} -\frac{5 l^2 }{v^2} )  \expval{T}  \nonumber \\ 
    &-\frac{6 D l^2}{v^4} .
    \label{eq:T3exp-sym} 
\end{align}
Notably, the exact expressions for the moments, Eqs.~(\ref{eq:Texp-sym}-\ref{eq:T3exp-sym}) can be obtained in few lines with the martingale approach, see Sec.~\ref{sec:ddd}.  The fact that  {\em all moments} of the exit time are written as functions of only its first moment $\langle T\rangle$ is a direct consequence of the fact that $T$ and $X_T$ are uncorrelated in the first-exit-time problem around a symmetric interval, see Sec.~\ref{sec:symmetry}.

\subsection{Two asymmetric absorbing boundaries}
\label{2-asymmetric}

Obtaining closed-form expression for the exit-time distribution $P(T)$ in the case of asymmetric absorbing boundaries about the origin is full of challenges.
Unlike the previous case with symmetric boundaries, where the exit site $X_T$ and exit time $T$ are statistically independent (see Sec.~\ref{sec:symmetry}), here it is not easy to make any such comment.
Consequently, the factorization in Eq.~\eqref{eq:27} that assumes the independence of $X_T$ and $T$ may not be employed for asymmetric boundaries. 
However, as discussed in Sec~\ref{sec:fes-asymm}, using the martingale approach we may obtain a closed-form expression for the mean exit time $\expval{T}$, see Eq.~\eqref{eq:21}. The higher moments $\expval{T^n}$ ($n>1$) of the exit time may be constructed from the mean $\expval{T}$, see Supplemental Material Sec.~\ref{Sup:Moments}. 

Despite the difficulty in obtaining $P(T)$, we remark that other non-trivial quantities can be effectively computed for the asymmetric boundaries employing the martingale approach. In this regard, a few examples are demonstrated in the next section.

\section{Area swept till the escape}
\label{sec:area}
Different from conventional techniques, we offer a direct approach of the problem of accessing  the statistics of the stochastic area. We employ Dynkin's Martingales which enable an extension of the problem beyond a single absorbing boundary and the first-escape time as it has been treated in previous works. In particular, by setting $g_t(x) = x^2$ in Eq.~\eqref{eq:Dynkin's Langevin}, we may write
\begin{eqnarray}
     \mathcal{Y}_t [g] &=& X_t^2 - X_0^2  - 2v \int_0^t \dd s \, X_s - 2D \int_0^t \dd s \nonumber\\
    &=& X_t^2 - X_0^2  - 2 v A_t -2Dt,
    \label{eq:Mt}
\end{eqnarray}
where $ A_t = \int_0^t \dd s \, X_s$ is the stochastic area swept by  $X_t$  in the time interval $[0,t]$, as defined in Eq.~\eqref{eq:area}. Applying Doob's OSTm to the martingale~\eqref{eq:Mt}, we obtain that  the mean area swept by the process till any stopping time $T$ obeying conditions I and II (see Sec.~\ref{sec:stop}) reads
\begin{equation}
   \langle A_T \rangle  
   =  \frac{\langle X_T^2 \rangle -  \langle X_0^2\rangle  - 2D\langle T \rangle}{2v}.
   \label{eq:meanarea}
\end{equation}
Equation~\eqref{eq:meanarea} holds for any stopping time obeying Doob's OSTm requirements, for any given initial density $\rho_0(X_0)$, and for all values of $v>0$.

In the following, we specialize our result~\eqref{eq:meanarea} to $T$ given by the first-exit time from the interval $(-l_{-},l_{+})$
for a drift-diffusion process starting at $X_0 = 0$ with drift velocity $v>0$.
By replacing the corresponding expressions for $\langle X^2_T \rangle$ and $\langle T \rangle$ (see Supplemental Material Sec.~\ref{Sup:Moments}) in Eq.~\eqref{eq:meanarea}, we obtain the formidable analytical expression
\begin{widetext}
\begin{equation}
\label{eq:mean_area_asymmetric}
\expval{A_T} =\left(\frac{D^2}{v^3}\right)\frac{\exp(\mathrm{Pe}^{+}) \left[\exp(\mathrm{Pe}^{-}) -1\right] \left[(\mathrm{Pe}^{+})^2 -2\mathrm{Pe}^{+}\right]  + \left[\exp(\mathrm{Pe}^{+}) -1\right] \left[(\mathrm{Pe}^{-})^2 +2\mathrm{Pe}^{-}\right]}{2 \left[\exp(\mathrm{Pe}^{+}+\mathrm{Pe}^{-})-1\right]},
\end{equation}
\end{widetext}
where 
\begin{equation}
    \mathrm{Pe}^- = \frac{v l_{-}}{D}, \quad\text{and}\quad\mathrm{Pe}^+ =\frac{ v l_{+}}{D},
    \end{equation}
    are the two characteristic P\'eclet numbers of the biased diffusion in the asymmetric interval $[-l_-,l_+]$. Notably, the analytical expression~\eqref{eq:mean_area_asymmetric}  reveals that the mean area till the first-exit time may be put in the scaling form
\begin{equation}
   \langle A_T \rangle =\frac{D^2}{v^3}  \Psi_A (\mathrm{Pe}^+,\mathrm{Pe}^-), \label{eq:mean_area_asymmetricsc}
\end{equation} 
with   
\begin{align}
\label{eq:Psi_A}
  \Psi_A (x,y) & \nonumber \\
  &\hskip-40pt =\frac{\exp(x) (\exp(y) -1) (x^2 -2x) + (\exp(x) -1) ( y^2 +2y)}{2 \left(\exp(x+y)-1\right)}      
\end{align}
being the scaling function.
In the limit $y\to \infty$, we have $\lim_{y\to \infty} \Psi_A (x,y) = (x^2-2x)/2$.
For the case of symmetric boundaries $l_{-} = l_{+} = l$, Eq.~\eqref{eq:mean_area_asymmetric} simplifies to 
\begin{equation}
    \langle A_T \rangle = \frac{l^2}{2 v} - \frac{Dl}{v^2} \tanh  \qty( \frac{vl}{2 D} ).
    \label{eq:mean_area_simme}
\end{equation}
Moreover, one can also show that for  $X_0 > 0$, $D = 1/2$ and $v<0$,  till the first-passage time to the origin, which is equivalent to take $l_{-} \rightarrow 0$ and $l_{+} \rightarrow \infty$ yields $\expval{X^m_T}\rightarrow 0$ and then $\langle A_T \rangle  =  (X_0^2/2v) + (X_0/2v^2)$, which was derived in previous works~\cite{Kearney_2005, Abundo2013}.

\begin{figure}[!htbp]
\centering
\includegraphics[scale=1.2]{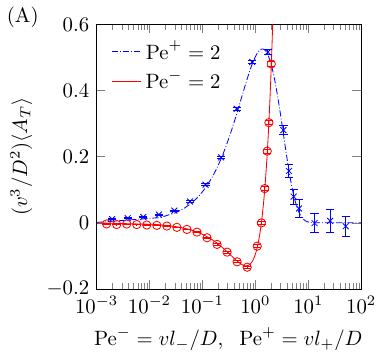}
\includegraphics[scale=1.2]{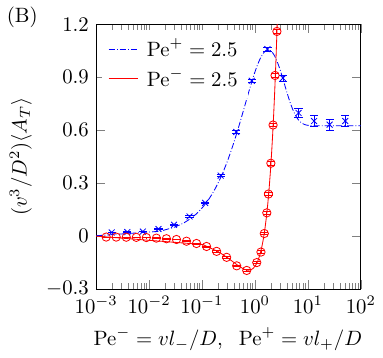}
\caption{Mean value of the  area $\langle A_T \rangle$ swept by a biased diffusion (scaled by $v^3/D^2$ and with initial position $X_0=0$) till its first-exit time from the interval $(-l_-,l_+)$ as a function of the characteristic P\'eclet numbers. Simulation results (symbols) are compared with analytical results given by Eq.~\eqref{eq:mean_area_asymmetric}  (lines). Red symbols and solid lines show the effect of increasing Pe$^+$ while keeping  Pe$^{-}$ constant (Pe$^{-} = 2$ in panel A and Pe$^{-} = 2.5$ in panel B) and  blue symbols and dash-dotted lines show the effect of increasing Pe$^-$ as keeping  Pe$^{+}$ constant (Pe$^{+} = 2$ in panel A and Pe$^{+} = 2.5$ in panel B). Note that in the asymptotic limit $v^3/D^2 \langle A_T \rangle$, tends to approach the value $\left[(\mathrm{Pe}^+)^2-2\mathrm{Pe}^+\right]/2$.
Error bars show the standard error of the mean. Simulation parameters: number of stochastic realizations $10^4$, simulation time step $10^{-4}$,  and $v =D=1$.   }
\label{fig:mean_area_analytic_simulation}
\end{figure}

\begin{figure}[!htbp]
\centering
\includegraphics[scale=0.45]{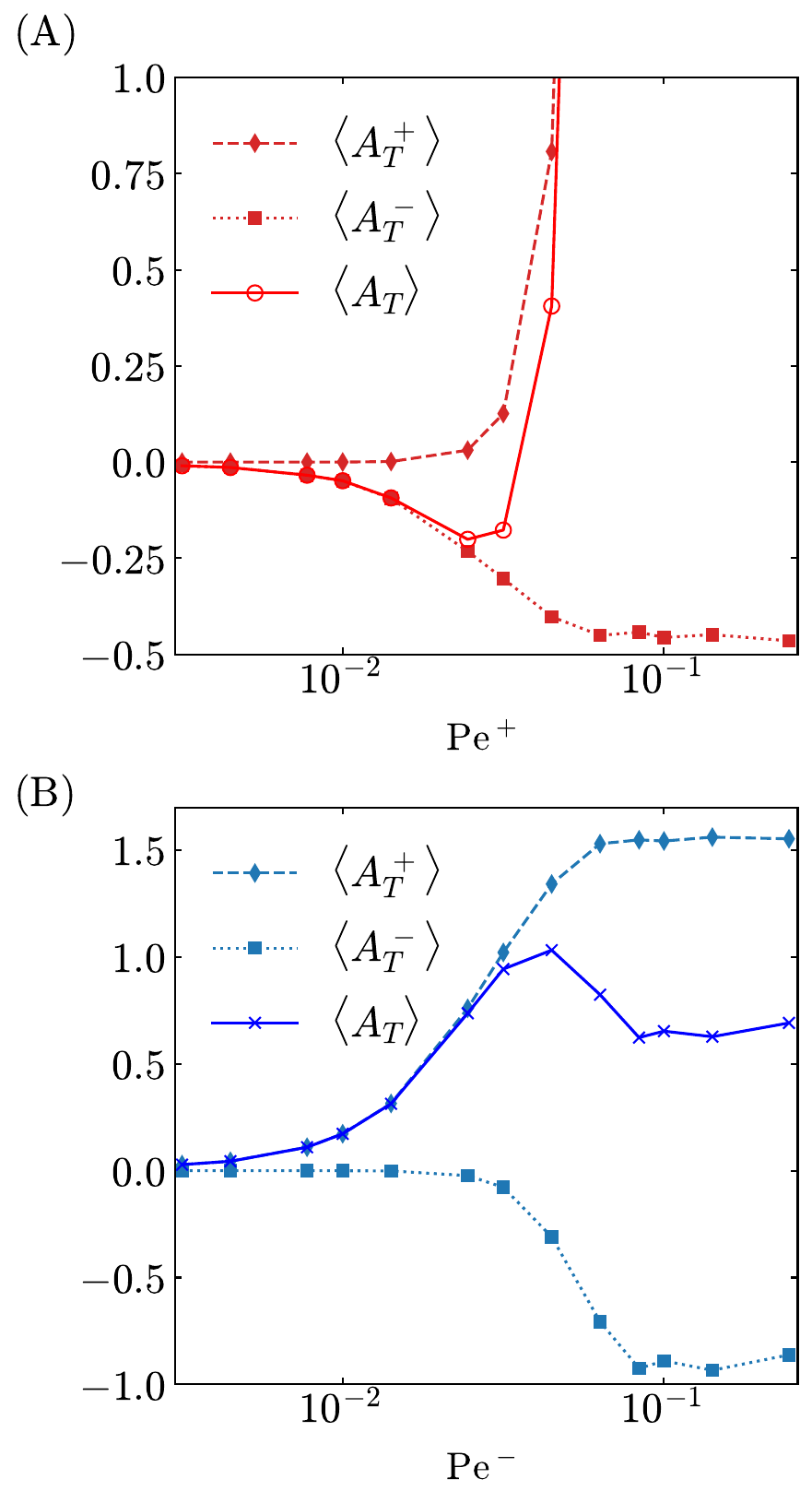}
\caption{Numerical estimates of $\langle A_T^+ \rangle$, $\langle A_T^-\rangle$ and $\langle A_T \rangle$ as a function of the characteristic P\'{e}clet numbers, obtained from numerical simulations. (A) Varying Pe$^+$ with  Pe$^-$ fixed at $2.5$. (B) Varying Pe$^-$ with  Pe$^+$ fixed at $2.5$. Symbols are obtained from numerical simulations, whereas the lines are guide to the eye. Simulation parameters: number of stochastic realizations $10^4$, simulation time step $10^{-4}$,  and $v =D=1$.} 
\label{fig:area_components}
\end{figure}

Figure~\ref{fig:mean_area_analytic_simulation} illustrates the mean value of the area  $\langle A_T \rangle$ (scaled by $v^3/D^2$) swept by the process $X_t$ until time $T$, highlighting  its dependence on the characteristic P\'eclet numbers. Simulation results (symbols) reproduce the analytical result Eq.~\eqref{eq:mean_area_asymmetric} (lines) with exquisite accuracy. 
Two distinct cases are explored:
the effect of increasing $l_+$ while maintaining $l_-$ fixed at the values of  Pe$^{-} = 2$ in panel A and Pe$^{-} = 2.5$ in panel B (red symbols and solid lines); and the effect of increasing $l_-$ while keeping $l_+$ fixed at the values of Pe$^{+} = 2$ in panel A and Pe$^{+} = 2.5$ in panel B (blue symbols and dash-dotted lines). 
First, we analyze what happens when varying Pe$^+$ at constant Pe$^-$. 
Interestingly, the mean area  decreases to negative values for smaller Pe$^+$ values, mainly because when $l_- \gg l_+$, negative contributions hold substantial weight.
However, with increasing Pe$^+$, $\langle A_T \rangle$ reaches a minimum  and positive contributions become more significant due to the dominance of the positive bias as we have $v>0$. This influence also leads to increasing the mean area with further increase in $\mathrm{Pe}^+$.
Such non-monotonicity observed in the mean area $\expval{A_T}$ arises due to the competing behavior of two components $A^{+}_T \equiv \int_0^T \dd{s} X_s \Theta(X_s)$ and $A^{-}_T \equiv \int_0^T \dd{s} X_s (1-\Theta(X_s))$, where $\Theta(x)$ represents the Heaviside function. The quantities $A^{+}_T$ and $A^{-}_T$ (referred to as positive and negative components of the area, respectively) denote the integrated area under the trajectory $X_{[0,T]}$ on the upper and lower halves of the axis $X_t = 0$, respectively (see Fig.~\ref{fig1}). Consequently, the total area is given by $A_T = A^{+}_T + A^{-}_T$ with $A^{+}_T \geq 0$ and $A^{-}_T \leq 0$.
When we increase Pe$^+$ at constant Pe$^-$, the positive component of the area keeps increasing while the negative one is exhausted reaching a saturation value. For smaller values of Pe$^{+}$ with fixed Pe$^{-}$, the mean area $\expval{A_T}$ is primarily influenced by the negative contribution of $\expval{A^-_T}$ as $\expval{A^+_T}$ remains negligible. With further increments in Pe$^+$, the positive contributions to $\expval{A_T}$ become significant as $\expval{A^+_T}$ increases rapidly due to positive drift, leading to the minimum of $\expval{A_T}$. For larger values of Pe$^+$, the average $\expval{A^-_T}$ reaches a saturation value due to fixed Pe$^{-}$, whereas $\expval{A^+_T}$ continues to rise because of  the positive drift. Numerical results illustrate the behavior of  $\expval{A^+_T}$, $\expval{A^-_T}$, and $\expval{A_T}$ with increasing Pe$^{+}$ at fixed Pe$^{-}$ in panel (A) of Fig.~\ref{fig:area_components}.

We now analyze the behavior of the scaled mean area with varying Pe$^-$ at constant Pe$^+$, i.e., the dash-dotted blue lines in Fig.~\ref{fig:mean_area_analytic_simulation}. The mean area   starts increasing for smaller Pe$^-$ values reaching a maximum. This trend again may be attributed to the interplay between the two distinct P\'eclet numbers in the system. For smaller values of Pe$^-$ with fixed Pe$^{+}$, $\expval{A^-_T}$ is negligible and the effect of the positive bias  initially  leads to an increase in $\expval{A_T}$ as the latter is dominated by $\expval{A^+_T}$. See panel (B) in Fig.~\ref{fig:area_components}. 
With further increments in Pe$^-$, trajectories that move against the bias sweep more space below the time axis, which increases the magnitude of $\expval{A^-_T}$ and adds negative contributions to $\expval{A_T}$.  
This negative contribution along with the positive contribution from $\expval{A^+_T}$ leads to the non-monotonicity of $\expval{A_T}$. For larger value of Pe$^-$, both $\expval{A^+_T}$ and $\expval{A^-_T}$ reach  saturation values (see panel (B) in Figure~\ref{fig:area_components}). The saturation of $\expval{A^+_T}$ occurs because of fixed Pe$^{+}$. On the other hand, the saturation of $\expval{A^-_T}$ arises because the trajectories eventually reach the positive threshold due to the large distance to the negative threshold. Finally, the scaled mean area for large Pe$^-$ with fixed Pe$^+$ saturates to the value $\left[(\mathrm{Pe}^+)^2-2\mathrm{Pe}^+\right]/2$.

\begin{figure}[!htbp]
\centering
\includegraphics[scale=0.45]{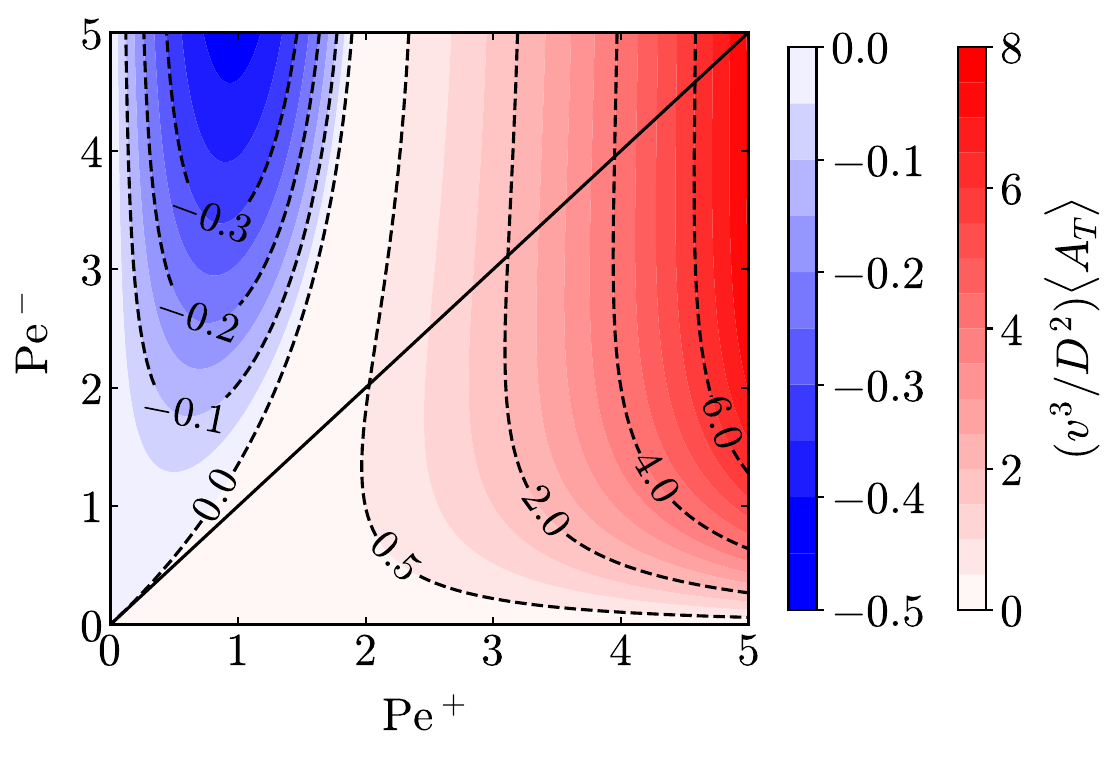}
\caption{Scaled mean area $(v^3/D^2) \langle A_T \rangle$ swept by a biased diffusion above the time axis till its first exit time from an interval as function of the characteristic P\'eclet numbers  $\mathrm{Pe}^+$ and $\mathrm{Pe}^-$. The colormap shows analytical results obtained from Eq.~\eqref{eq:mean_area_asymmetric}. The solid black line has slope one, representing the region of symmetric absorbing boundaries. The dashed lines correspond to isolines for different values of the scaled mean area.} 
\label{fig:mean_area_countour}
\end{figure}

Figure~\ref{fig:mean_area_countour} displays the analytical value  of the mean area $\langle A_T \rangle$  as a function of both Pe$^+$ and Pe$^-$, suggesting that the  mean area is strongly influenced by the bias. The colormap in  Fig.~\ref{fig:mean_area_countour}  illustrates the impact of asymmetric absorbing boundaries on the  mean area~$\expval{A_T}$ as given by Eq.~\eqref{eq:mean_area_asymmetric}. Interestingly, in the region where Pe$^+$ and Pe$^-$ are equal (indicated by the solid black line), the absorbing boundaries are symmetric, and the mean area is always positive, see Eq.~\eqref{eq:mean_area_simme}.  It is  also worth noticing that there exists a region (between the solid black line with slope one and the dashed black line depicting $\langle A_T \rangle = 0$), in which the positive threshold is closer than the negative one $l_+<l_-$ but yet the mean area is positive. We attribute this effect to the fact that the system is biased towards~$l_+$. This bias results in a greater accumulation of area in the positive region, and a correspondingly smaller area in the negative region. This difference is particularly pronounced at large values of Pe$^+$, where the area becomes significantly larger than the corresponding area at large Pe$^-$ values.

As a concluding note, we attack a hitherto unexplored quantity to our knowledge, by  computing the cross-correlation function $C(X_T,T)$  between the first-exit time $T$ and the exit site $X_T$.
To this aim, we apply Doob's OSTm to a wisely-chosen Dynkin's martingales, and make use of our analytical expression for the mean area, together with previous knowledge on first-passage theory~\cite{Goel1974}. Identifying the Dynkin martingale associated with the function  $g_t(x) = t x$ [see Eq.~\eqref{eq:Dynkin's Langevin}] as ${\mathcal{Y}}_t = t X_t - \int_0^t {\mathrm{d}} s (X_t + v t) = t X_t - A_t - v t^2/2$, and subsequently applying Doob's OSTm, i.e., $\expval{{\mathcal{Y}}_T} = \expval{{\mathcal{Y}}_0} = 0$, we obtain
\begin{equation}
    \langle  X_T T\rangle = \langle A_T \rangle + \frac{v \langle T^2 \rangle}{2},
    \label{eq:52}
\end{equation}
which holds for any stopping time that obeys conditions~I and II in Sec.~\ref{sec:stop}.
Equation~\eqref{eq:52} implies that
\begin{equation}
    \label{eq:XT_T_cross}
   \langle  X_T T\rangle = \frac{v}{2} \expval{T^2} -\frac{D\langle T \rangle}{v} +  \frac{\langle X_T^2\rangle -X_0^2 }{2v},
\end{equation}
where we have used Eq.~\eqref{eq:meanarea}. Note that Eq.~\eqref{eq:XT_T_cross} is the same as Eq.~\eqref{eq:MET2-gen}, which is obtained from the Dol\'{e}ans--Dade exponential family.
Using Eqs.~\eqref{eq:MET-gen-x0} and~\eqref{eq:XT_T_cross}, the cross-correlation function $C(X_T,T) =\expval{X_T T}-\expval{X_T}\expval{T}$ may then be written as
\begin{align}
 \label{eq:CXTT0}
     C(X_T,T) &= \frac{v}{2} \left(\expval{T^2} -2 \expval{T}^2 \right)- \qty( X_0+ \frac{D}{v}) \expval{T} \nonumber \\
     &+ \frac{\expval{X_T^2} - X_0^2}{2v},
 \end{align}
which depends on the first and second moments of $T$ and the second moment of $X_T$.

In particular, for the case of two absorbing boundaries positioned asymmetrically around the origin,  we may use the corresponding expressions for $\expval{T}, \expval{T^2}$, and $\expval{X_T^2}$ in Eq.~\eqref{eq:CXTT0} (see Supplemental Material Sec.~\ref{Sup:Moments}) and obtain
\begin{widetext}
\begin{align}
      \label{eq:corr_asyme}
C(X_T, T) = \left( \frac{D^2}{v^3}\right) \frac{ \exp(\mathrm{Pe}^{+}) (\mathrm{Pe}^{-}+\mathrm{Pe}^{+})}{ \left\{  \exp\left(\mathrm{Pe}^+ + \mathrm{Pe}^- \right) - 1 \right\}^2} & \Big[ \mathrm{Pe}^+ \left\{ \exp\left(\mathrm{Pe}^-\right)-1\right\}  \left\{\exp\left(\mathrm{Pe}^+\right)+1\right\}  \nonumber \\ 
&- \mathrm{Pe}^- \left\{\exp\left(\mathrm{Pe}^-\right) +1\right\} \left\{ \exp\left(\mathrm{Pe}^+\right)-1\right\} \Big].
\end{align}
\end{widetext}
Equation~\eqref{eq:corr_asyme} may be put in the scaling form
\begin{align}
\label{eq:C(X_T, T)_in_Psi_C}
C(X_T,T) = \frac{D^2}{v^3}  \Psi_C (\mathrm{Pe}^+,\mathrm{Pe}^-),
\end{align}
with the scaling function in this case given by
\begin{align}
\label{eq:Scaling_corr}
\Psi_C(x,y) &=  \dfrac{\exp(x)(x+y)}{\left[\exp(x+y)-1\right]^2}   \Big[ (\exp(y)-1 ) (\exp(x)+1) x  \nonumber \\
& - (\exp(y)+1)(\exp(x)-1)y \Big].
\end{align}

\begin{figure}[!htbp]
\centering
\includegraphics[scale=0.45]{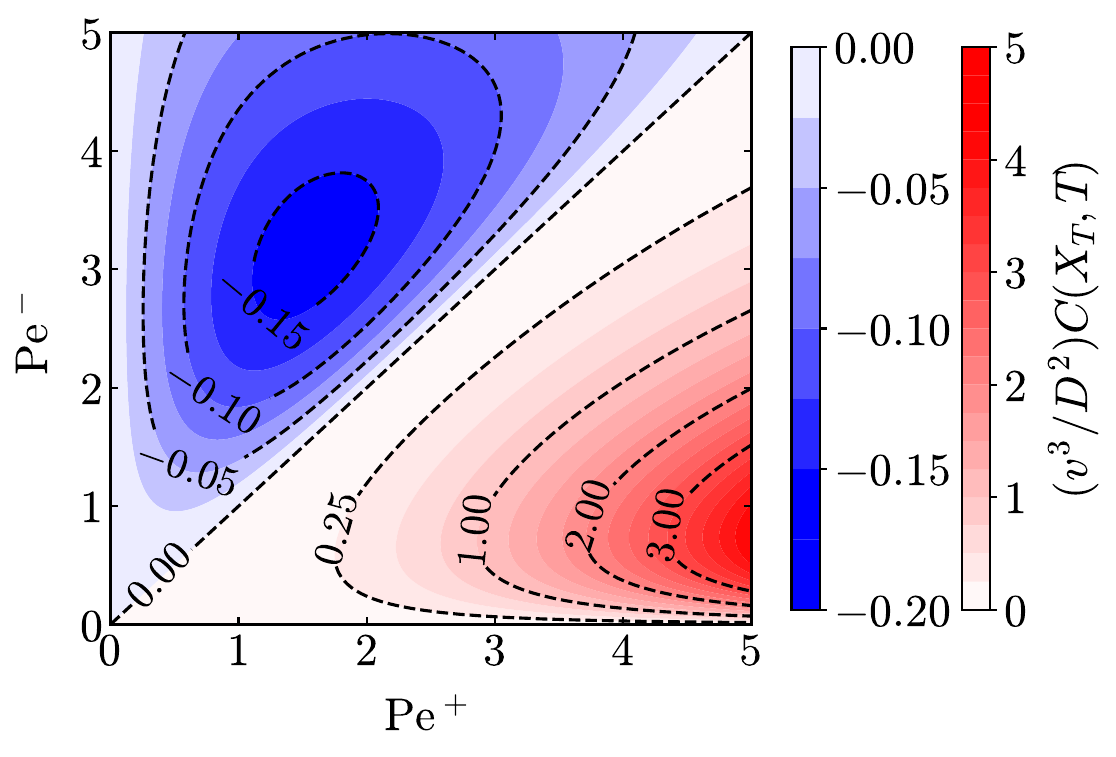}
\caption{Scaled cross-correlation function $(v^3/D^2)  C(X_T, T)$ between the first-exit site $X_T$ and the first-exit time $T$ as function of the characteristic P\'eclet numbers  $\mathrm{Pe}^+$ and $\mathrm{Pe}^-$. The colormap shows analytical results obtained from Eq.~\eqref{eq:C(X_T, T)_in_Psi_C}. Dashed lines represent isolines for different values of the scaled cross-correlation function. Note that the isoline corresponding to $(v^3/D^2)  C(X_T, T) =0$, where $X_T$ and $T$ are uncorrelated, has slope one representing the region of symmetric absorbing boundaries.} 
\label{fig:cross_corr}
\end{figure}
Figure~\ref{fig:cross_corr} shows the analytical value of the scaled cross-correlation function $({v^3}/{D^2})\,C(X_T, T)$ [Eq.~\eqref{eq:corr_asyme}] as a function of  the characteristic P\'eclet numbers $\mathrm{Pe}^+$ and $\mathrm{Pe}^-$. The black line with a slope of one represent the region associated with symmetric absorbing boundaries, indicating that only within this region $X_T$ and $T$ are   uncorrelated, i.e. $C(X_T, T) = 0$. This result is in agreement with the cross-correlation symmetry conjectured  in Sec.~\ref{sec:symmetry}. Unlike for the mean area $\langle A_T\rangle$, the cross-correlation function  $C(X_T, T)$ is symmetric around the region $\mathrm{Pe}^+=\mathrm{Pe}^-$, with the latter being a separatrix between two distinct regimes. When  $\mathrm{Pe}^+ > \mathrm{Pe}^-$, there exists a positive correlation between the exit site and the exit time, whereas when  $\mathrm{Pe}^- > \mathrm{Pe}^+$ the exit site and the exit time are anticorrelated. From this analysis and our exact formula~\eqref{eq:corr_asyme} we conclude that positive $X_T$--$T$ cross-correlations  take place in asymmetric escape problems with the largest threshold in the direction of the drift, whereas  negative  $X_T$--$T$ cross-correlations  appear  when the the largest threshold is in  the direction opposite to the net drift. Such analysis could be of particular interest in the study of binary decision making in biology and neuroscience~\cite{siggia2013decisions,wang2015landscape,de2012protein,durmaz2023human}.  In particular,  pushing forward our theory to more complex dynamics may result in inference schemes for the decision thresholds based on measurements of the cross-correlators $C(X_T,T)$.

\section{Conclusions}
\label{sec:conclu}
Our work delves into the statistical analysis of one-dimensional biased diffusion governed by the overdamped Langevin equation. Specifically, we have focused on the dynamics within an interval and investigated first-exit-time statistics, shedding light on various key aspects of the process. We have addressed the probability density $P(T)$ for the first exit time $T$ from an interval. We have also examined the splitting probabilities for the particle to exit from the interval and the correlation between the first-exit time and the corresponding first-exit position. Incorporating tools from martingale theory, our approach enables the derivation of analytical results in first-passage theory. Notably, our calculations yield concise outcomes compared to traditional methods, such as path integrals and Fokker-Planck equations. Moreover, we have extended our analysis to the stochastic area swept by the biased diffusion for any class of stopping time, encompassing first-exit time as a specific example, contributing to the generalization of previous results in the literature. Finally, we have established general relations between moments and cross-correlations for any class of stopping times. In essence, our work not only presents a comprehensive study of biased diffusion dynamics within bounded intervals but also offers analytical shortcuts and broad insights into the statistical properties of first-exit times, positions, and associated stochastic areas. The utilization of martingale theory becomes a powerful tool in unraveling the intricacies of these stochastic processes, opening new doors for better understanding and further advancements of diffusion phenomena.

Besides statistical physics and decision theory, our results could have impact even in picturesque fields such as game theory and gambling. Suppose a gambler bets on the first-exit site of a drifted Brownian particle with negative drift $v<0$, with its wealth in each game being to exit site value $X_T$. Then, no matter which threshold values $l_+$ and $l_-$ he/she chooses, the game will lead to a net loss $\langle X_T\rangle\leq 0$. If however, the gambler's prize is based upon the area (i.e. the  value of the time-integrated value of the particle position) then it is possible to find gambling strategies such that the net wealth $\langle A_T\rangle>0$. Such strategies may be borrowed from the results in Sec.~\ref{sec:area} by just repeating  the calculations but for the case $v<0$.
Even if reminiscent of Parrondo's paradoxical games~\cite{parrondo2000new}, here the winning strategies do not require the switch between two strategies but rather just quitting the game at wisely-chosen threshold values $l_+$ and $l_-$. Furthermore, we expect such {\em winning} stopping strategies to inspire energy-extraction  protocols in stochastic thermodynamics~\cite{neri2019integral,manzano2021thermodynamics,yang2023fluctuation}.

\clearpage
\onecolumngrid
\beginsupplement

\section*{Supplemental Material}
\label{sec:sm}

\setcounter{section}{0}
\setcounter{equation}{0}
\setcounter{figure}{0}

\section{Moments of the first-exit time and the first-exit site from an asymmetric interval or arbitrary length}
\label{Sup:Moments}

We examine the case  of two absorbing boundaries located at $-l_{-}$ and $l_{+}$ with $l_{\pm} \geq 0$ for the process $X_t$ given in Eq.~\eqref{eq:eom_Xit}. Here, the stopping time is given by $T = \text{inf} \qty{ t \geq 0 | X_t \notin [-l_{-}, l_{+}] }$ with $X_0 = 0$. Henceforth, we will employ the notation, $\exp(x) = \ee^{x}$, in order to ensure clarity. The first moment of the stopping time $T$ is given by Eq.~\eqref{eq:21}$, \expval{T} = \qty(l_{+} P_{+}  - l_{-} P_{-})/v $. By substituting Eq.~\eqref{eq:P+P-assym_FB_x0} in Eq.~\eqref{eq:21}, we obtain the first moment of the exit time as
\begin{equation}
\label{eq:first_moment_asymm}
  \expval{T} =  \frac{D \left[\mathrm{Pe}^- + \ee^{\mathrm{Pe}^+} \left(\ee^{\mathrm{Pe}^-} \mathrm{Pe}^+ - \mathrm{Pe}^- - \mathrm{Pe}^+\right)\right]}{v^2 \left(\ee^{\mathrm{Pe}^- + \mathrm{Pe}^+} - 1\right)} .
\end{equation}
Equation~\eqref{eq:first_moment_asymm} may be also obtained using the relation~\cite{Goel1974}
\begin{equation}
\label{eq:s2}
    \expval{T}(X_0) =   \frac{1}{D} \left[ \int_{X_0}^{l_+} \dd \eta \, \pi(\eta) \int_{-l_-}^{\eta} \left(  \pi(\xi)\right)^{-1} \dd \xi - R(X_0) \int_{-l_-}^{l_+} \dd \eta \, \pi(\eta) \int_{-l_-}^{\eta} \left(\pi(\xi)\right)^{-1} \dd \xi\right],
\end{equation}
where $\expval{T}(X_0)$ denotes the mean exit time as a function of the initial position $X_0$, and we have $\pi(y) \equiv \exp(-v y /D )$ and 
$R (X_0) \equiv \left(\exp(\mathrm{Pe}^+ - v X_0/D) -1\right)  /\left(\exp(\mathrm{Pe}^+ +\mathrm{Pe}^-) -1\right)$.
In general, the $n-$th moment of the exit time  may be written as~\cite{Goel1974}
\begin{equation}
\label{eq:s3}
    \expval{T^n}(X_0) =  \frac{n}{D}\left[ \int_{X_0}^{l_+} \dd \eta \, \pi(\eta) \int_{-l_-}^{\eta} \frac{\expval{T^{n-1}}(\xi)}{ \pi(\xi)}   \dd \xi - R(X_0) \int_{-l_-}^{l_+} \dd \eta \, \pi(\eta) \int_{-l_-}^{\eta} \frac{\expval{T^{n-1}}(\xi)}{ \pi(\xi)}   \dd \xi \right] .
\end{equation}
In particular, for $X_0 =0$, the second moment of the exit time obtained by substituting Eq.~\eqref{eq:s2} in Eq.~\eqref{eq:s3} reads
\begin{align}
\expval{T^2} = \frac{D^2}{v^4 \left(\ee^{\mathrm{Pe}^{-}+\mathrm{Pe}^{+}}-1\right)^2}  &  \Big[ \left(\ee^{\mathrm{Pe}^{-}}-1\right) \ee^{\mathrm{Pe}^{+}} (\mathrm{Pe}^{+})^2 \left(\ee^{\mathrm{Pe}^{-}+\mathrm{Pe}^{+}}+3\right)  
-(\mathrm{Pe}^{-})^2 \left(\ee^{\mathrm{Pe}^{+}}-1\right) \left(3 \ee^{\mathrm{Pe}^{-}+\mathrm{Pe}^{+}}+1\right)  \nonumber \\
&+2 \left(\ee^{\mathrm{Pe}^{-}+\mathrm{Pe}^{+}}-1\right) \left\{\mathrm{Pe}^{-}-\ee^{\mathrm{Pe}^{+}} \left(-\ee^{\mathrm{Pe}^{-}} \mathrm{Pe}^{+}+\mathrm{Pe}^{-}+\mathrm{Pe}^{+}\right)\right\}
  \nonumber \\
& -4 \mathrm{Pe}^{-} \mathrm{Pe}^{+} \left(-2 \ee^{\mathrm{Pe}^{-}+\mathrm{Pe}^{+}}+\ee^{\mathrm{Pe}^{-}+2 \mathrm{Pe}^{+}}+\ee^{\mathrm{Pe}^{+}}\right)  \Big]  .  
\end{align}

The $m$-th moment of $X_T$ is given by Eq.~\eqref{eq:20}, $\expval{X^m_T} =  (- l_{-})^m P_{-} + l^m_+ P_{+}$. Using  Eq.~\eqref{eq:P+P-assym_FB_x0} in the latter expression, we retrieve  the first two moments of the exit site for the case $X_0=0$
\begin{equation}
   \expval{X_T} = \frac{D \left[\mathrm{Pe}^{-} + \ee^{\mathrm{Pe}^{+}} \left(\ee^{\mathrm{Pe}^{-}} \mathrm{Pe}^{+} - \mathrm{Pe}^{-} - \mathrm{Pe}^{+}\right)\right]}{v \left(\ee^{\mathrm{Pe}^{-} + \mathrm{Pe}^{+}} - 1\right)},
\end{equation}
\begin{equation}
    \expval{X_T^2} = \frac{D^2 \left[ \left(\ee^{\mathrm{Pe}^{+}} - 1\right)(\mathrm{Pe}^{-})^2 + \left(\ee^{\mathrm{Pe}^{-}} - 1\right) \ee^{\mathrm{Pe}^{+}} (\mathrm{Pe}^{+})^2\right]}{v^2 \left(\ee^{\mathrm{Pe}^{-} + \mathrm{Pe}^{+}} - 1\right)}.
\end{equation}


\bibliographystyle{unsrt}
\bibliography{refs}

\end{document}